\documentclass{llncs}

\usepackage{times}  
\usepackage[noend]{algpseudocode}
\usepackage{algorithm}
\usepackage{textcomp}
\usepackage{lineno}
\usepackage{subfig}
\usepackage{graphicx}
\usepackage[utf8]{inputenc}
\usepackage{todonotes}
\usepackage{array}
\usepackage{amsmath}
\usepackage{amssymb}

\newcolumntype{P}[1]{>{\centering\arraybackslash}p{#1}}
\newcolumntype{M}[1]{>{\centering\arraybackslash}m{#1}}
\renewcommand{\algorithmicrequire}{\textbf{Input: A link stream $L$ defined over a set of nodes $V$}}
\renewcommand{\algorithmicensure}{\textbf{Output: The set of all maximal cliques in $L$}}
\newcommand{\noteperso}[1]{}
\newcommand{\ie}{\emph{i.e.}}

\newcommand{\includegraphicsmaybe}[2][]{\IfFileExists{#2}{\includegraphics{#2}}{\fbox{#2 is missing}\\}}

\begin{document}


\title{Discovering Patterns of Interest in IP Traffic\\ Using Cliques in Bipartite Link Streams}
\author{Tiphaine Viard\textsuperscript{1}, Raphaël
  Fournier-S'niehotta\textsuperscript{1},\\ Clémence Magnien\textsuperscript{2}, Matthieu Latapy\textsuperscript{2}}

\institute{\textsuperscript{1}\footnotesize CEDRIC CNAM, F-75003 Paris, France \\ tiphaine.viard@cnam.fr, fournier@cnam.fr\\
\textsuperscript{2}\footnotesize Sorbonne Universités, UPMC Univ Paris 06, CNRS, UMR 7606, LIP6, F-75005, Paris, France \\ clemence.magnien@lip6.fr, matthieu.latapy@lip6.fr}

\maketitle

\begin{abstract}

Studying IP traffic is crucial for many applications. We focus here on the detection of (structurally and temporally) dense sequences of interactions, that may indicate botnets or coordinated network scans. More precisely, we model a MAWI capture of IP traffic as a link streams, \ie{} a sequence of interactions $(t_1,t_2,u,v)$ meaning that devices $u$ and $v$ exchanged packets from time $t_1$ to time $t_2$. This traffic is captured on a single router and so has a bipartite structure: links occur only between nodes in two disjoint sets. We design a method for finding interesting bipartite cliques in such link streams, i.e. two sets of nodes and a time interval such that all nodes in the first set are linked to all nodes in the second set throughout the time interval. We then explore the bipartite cliques present in the considered trace. Comparison with the MAWILab classification of anomalous IP addresses shows that the found cliques succeed in detecting anomalous network activity.

\end{abstract}

\section{Introduction}

Attacks against online services, networks, and information systems, as well as identity thefts, have annual costs in billions of euros.
Network traffic analysis and anomaly detection systems are of crucial help in fighting such attacks.
In particular, much work is devoted to the detection of anomalous patterns in traffic. 
This work mostly relies on pattern search in (sequences of) graphs, which poorly captures the dynamics of traffic, or on signal analysis, which poorly captures its structure. Instead, we use here the {\em link stream} framework, which allows us to model both structural and temporal aspects of traffic in a consistent way. Within this framework, a clique is defined by a set of nodes and a time interval such that all the nodes in that set continuously interact with each other during this time interval. Such patterns may be the signature of various events of interest like botnets, DDoS, and others.

We present in Section~\ref{sec-model} our {\em link stream} modeling of traffic captures, which calls for specific notions, as this traffic is bipartite. We present in Section~\ref{sec-finding} fast heuristics for finding cliques of interest in practice. We briefly present the MAWI dataset that we use, then detail the results of our experiments in Section~\ref{sec-results}. Related work is overviewed in Section~\ref{sec-rw}, and we discuss the main perspectives in Section~\ref{sec-conclusion}.

\section{Traffic as a Link Stream}
\label{sec-model}

IP traffic is composed of packets, each with a source address and a destination address. Routers forward these packets towards their destination, and one may capture traffic traces by recording the headers of packets managed by a router, along with the time at which they manage them. Such traces are generally collected on firewalls, access points, or border routers. As a consequence, they often have a bipartite nature: they capture exchanges between two disjoint sets of devices (for instance the ones in a company LAN and the outside internet), as these routers are not in charge of traffic between devices within one of these sets. This leads to the definition of {\bf bipartite link streams}, that extends the classical definitions of bipartite graphs \cite{DBLP:journals/socnet/LatapyMV08} and of link streams \cite{1710.04073,DBLP:journals/tcs/ViardLM16}.

\begin{figure}[!h]
\centering
\includegraphics[width=.5\columnwidth]{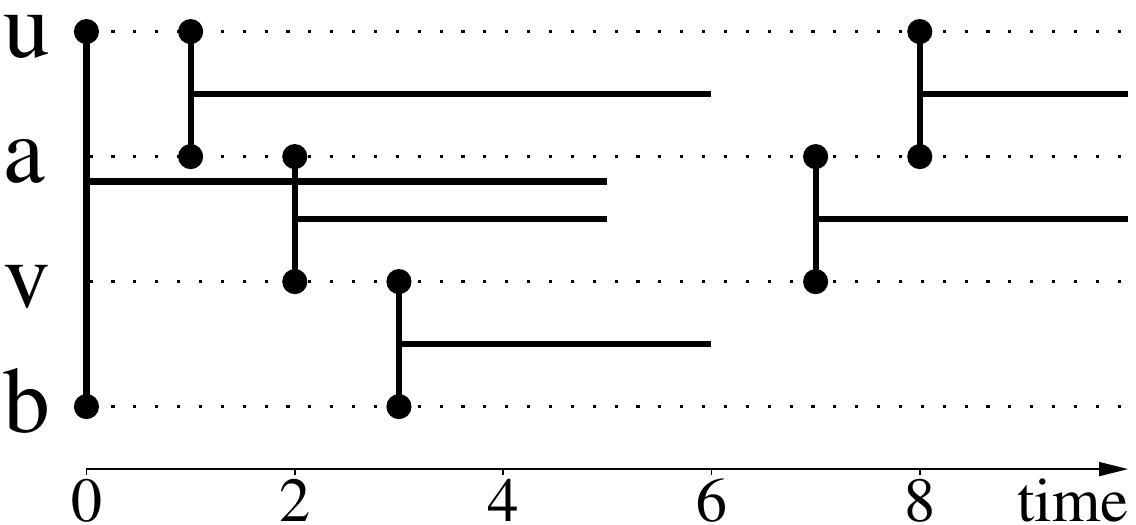}
\caption{
A bipartite link stream $L = (T,\top,\bot,E)$ with $T=[0,10]$, $\top=\{u,v\}$, $\bot=\{a,b\}$, and $E = ([1,6]\cup[8,10])\times\{(u,a)\} \cup [0,5]\times\{(u,b)\} \cup ([2,5]\cup[7,10]\times\{(v,a)\} \cup [3,6]\times\{(v,b)\}$. In other words, the links of $L$ are $(1,6,u,a)$, $(8,10,u,a)$, $(0,5,u,b)$, $(2,5,v,a)$, $(7,10,v,a)$, and $(3,6,v,b)$. We display nodes vertically and time horizontally, each link being represented by a vertical line at its beginning that indicate its extremities, and an horizontal line that represents its duration.
}
\label{fig:linkstream}
\end{figure}

A bipartite link stream $L = (T,\top,\bot,E)$ is defined by a time span $T$, a set of top nodes $\top$, a set of bottom nodes $\bot$, and a set of links $E \subseteq T \times \top \times \bot$. If $(t,u,v) \in E$ then we say that $u$ and $v$ are linked at time $t$. We say that $l = (b,e,u,v)$ is a link of $L$ if $[b,e]$ is a maximal interval of $T$ such that $u$ and $v$ are linked at all $t$ in $[b,e]$. We call $\overline{l} = b-e$ the duration of $l$. See Figure~\ref{fig:linkstream} for an illustration. We consider undirected links only: we make no distinction between $(t,u,v)$ and $(t,v,u)$ in $E$.


To model traffic as a link stream, we transform packets exchanged at timestamps into continuous interactions; we consider that two IP addresses are continuously linked together from time $b$ to time $e$ if they exchange at least one packet every second within this time interval. This leads to the following definition of $E$: it is the set of all $(t,u,v)$ such that $u$ and $v$ exchanged a packet at a time $t'$ such that $|t'-t|$ is lower than half a second. 


\section{Finding Cliques of Interest}
\label{sec-finding}

In a bipartite graph $G = (\top, \bot ,E)$, a clique is a couple $(C_\top, C_\bot)$ with $C_\top \subseteq \top$ and $C_\bot \subseteq \bot$ such that $C_\top\times C_\bot \subseteq E$. In other words, there is a link between each element of $C_\top$ and each element of $C_\bot$.

\begin{figure}[!h]
    \centering
    \ \hfill
    \includegraphics[width=.40\columnwidth]{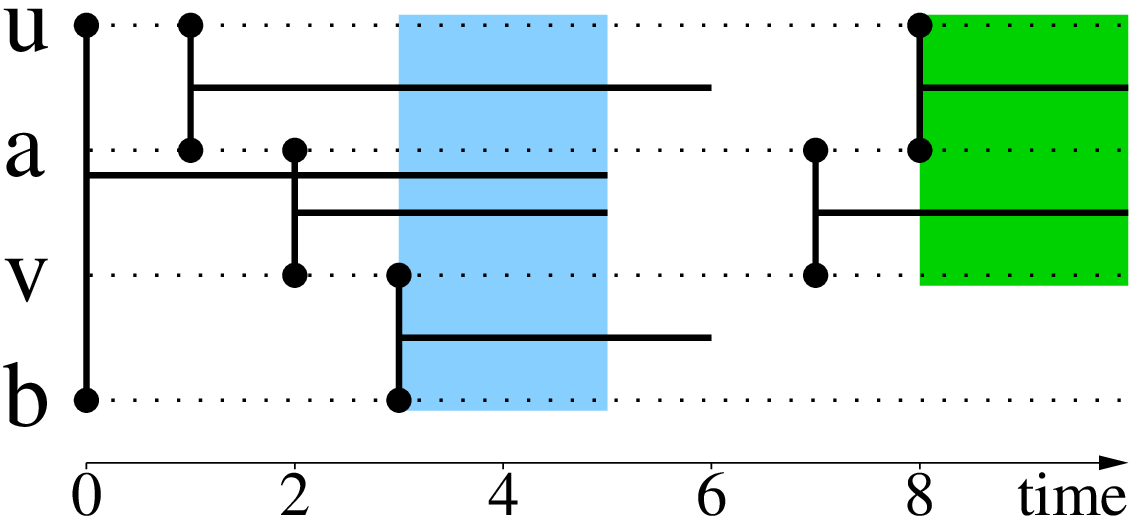}
    \hfill
    \includegraphics[width=.40\columnwidth]{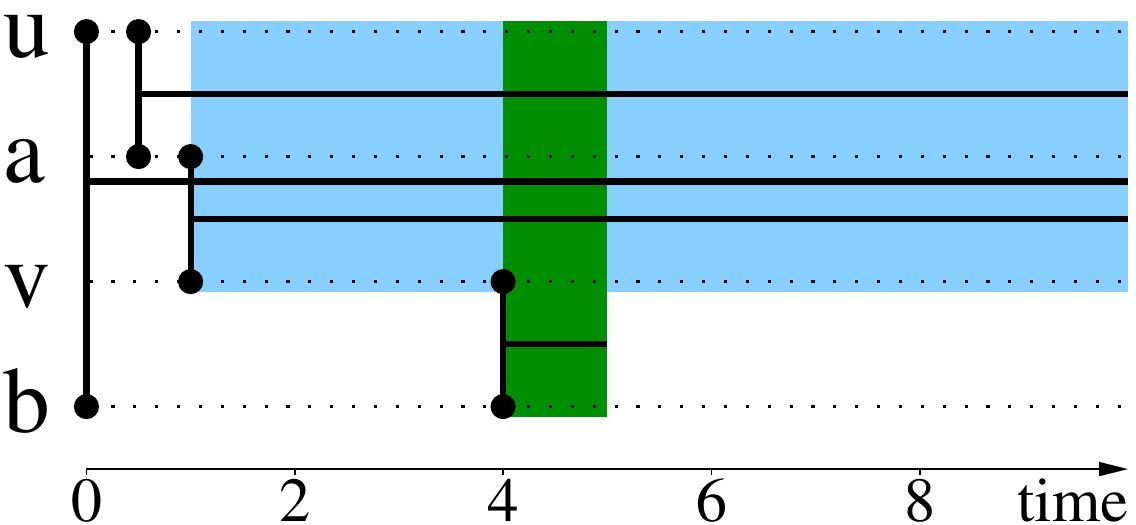}
    \hfill\
    \caption{{\bf Left:} Two maximal cliques of the bipartite link stream of Figure~\ref{fig:linkstream}: $(\{u,v\}, \{a,b\}, [3,5])$ (in blue) and $(\{u,v\},\{a\}, [8,10])$ (in green). {\bf Right:} From the blue clique $(\{u,v\},\{a\}, [1,10])$ our algorithm builds the clique $(\{u,v\},\{a,b\}, [4,5])$ by adding $b$, which reduces the time span from $[1,10]$ to $[4,5]$.}
    \label{fig:cliqueexemple}
\end{figure}

We define similarly a clique in a bipartite link stream as a tuple $(C_\top, C_\bot, I)$ with $C_\top \subseteq \top$, $C_\bot \subseteq \bot$ and $I$ an interval of $T$ such that $I\times C_\top\times C_\bot \subseteq E$. In other words, each element of $C_\top$ is linked to each element of $C_\bot$ for the whole duration of $I$. We call $|C_\top \cup C_\bot|$ the size of the clique and $|I|$ its duration. A clique is maximal if there is no other clique $(C'_\top, C'_\bot, I')$ such that $C_\top \subseteq C'_\top$, $C_\bot \subseteq C'_\bot$ and $I\subseteq I'$. See Figure~\ref{fig:cliqueexemple} for an illustration.

We explore cliques as patterns of interest in IP traffic. It is easy to extend to the bipartite case the (non-bipartite) algorithm presented in \cite{DBLP:conf/asunam/HimmelMNS16}. However, we face two issues. First, clique computations are costly, and enumerating all cliques is out of reach in our case; we therefore resort to sampling. Second, not all maximal cliques are interesting: 
in particular, a node and its neighbors is a bipartite clique, but has little interest for us.

As a consequence, we focus on {\em balanced} cliques: a clique $(C_\top, C_\bot, I)$ is balanced if and only if $| |C_\top|-|C_\bot| | \leq 1$. We then search for balanced cliques as follows. We iteratively build $(C_\top, C_\bot, I)$ from the empty clique of maximal duration, $(\emptyset, \emptyset, T)$. At each step we choose a random node $v$ in $\top$ or $\bot$ such that $v$ is linked to all nodes in $C_\bot$ or $C_\top$ respectively, during an interval $I_v \subseteq I$, and we update the current clique into $(C_\top \cup \{v\}, C_\bot, I_v)$ or $(C_\top, C_\bot \cup \{v\}, I_v)$, respectively. In order to ensure the obtained clique is balanced, we alternatively choose $v$ in $\top$ and in $\bot$ at each iteration. We stop when no node can be added.

This process samples a maximal balanced clique, and we run it many times to obtain a large set of such cliques (see Section~\ref{sec-results}). It is clear that this set may be biased by the sampling process, which is not uniform. However, since our primary goal is to explore the relevance of {\em some} cliques in the context of IP traffic, studying this bias is out of our scope. Notice also that the found clique is maximal amongst the set of balanced cliques, but not necessarily maximal in the stream: it may be included in an unbalanced clique.

Our sampling process tries to add as many nodes to the clique as possible, which generally induces a reduction of its duration, see Figure~\ref{fig:cliqueexemple}. As we are both interested in large {\em and} long balanced cliques, we include in the sampled set all the intermediary maximal balanced cliques built during the process.

\section{Dataset and Results}
\label{sec-mawi}

We use an IP traffic capture from the MAWI archive, more precisely from the DITL~\footnote{\url{http://mawi.wide.ad.jp/mawi/ditl/ditl2013/}} initiative, from 2013, June, 24\textsuperscript{th}, 23:45 to 2013, June 25\textsuperscript{th}, 00:45. This trace lasts $3,600$ seconds, during which $88,266,534$ packets are sent involving $992,466$ nodes. $408,751$ nodes are part of WIDE, and $583,715$ are outside of WIDE. WIDE/Non-WIDE sets are nearly balanced, improving our chances of finding large balanced cliques. 

We transform the data into a bipartite link stream $L=(T, \top, \bot, E)$, where $T= [0,3600]$, $\top$ contains all observed WIDE IP addresses, and $\bot$ all other observed IP addresses. As said in Section~\ref{sec-model}, $E$ is the set of links obtained by considering that nodes interact for one second every time they exchange a packet. $E$ contains $6,206,295$ links.



In addition to this raw data, we use the MAWILab database~\cite{mawilab}, which gives labels locating traffic anomalies in the MAWI archive. These labels are obtained using an advanced graph-based methodology that compares and combines different and independent anomaly detectors. This database indicates that there is a total of $488$ anomalous IP addresses in our dataset, that we will use to interpret the results in the following.

\label{sec-results}

\bigskip
We ran our sampling algorithm on the MAWI dataset. We ran 14 instances in parallel on a server\,\footnote{A linux machine with 24 cores at 2.9 GHz and 256 GB of RAM.}, for 106 days (3 months and 17 days). This led to a total of 1,291,084,661 sampled cliques, among which there were a great number of duplicates: $198,718,323$ are distinct maximal balanced cliques, which we study in the rest of the paper.

We call {\em anomalous clique} a clique which contains at least one IP address that is flagged as anomalous in the MAWILab database.

\begin{figure}[!h]
\centering
\ \hfill
\includegraphics[width=0.45\columnwidth]{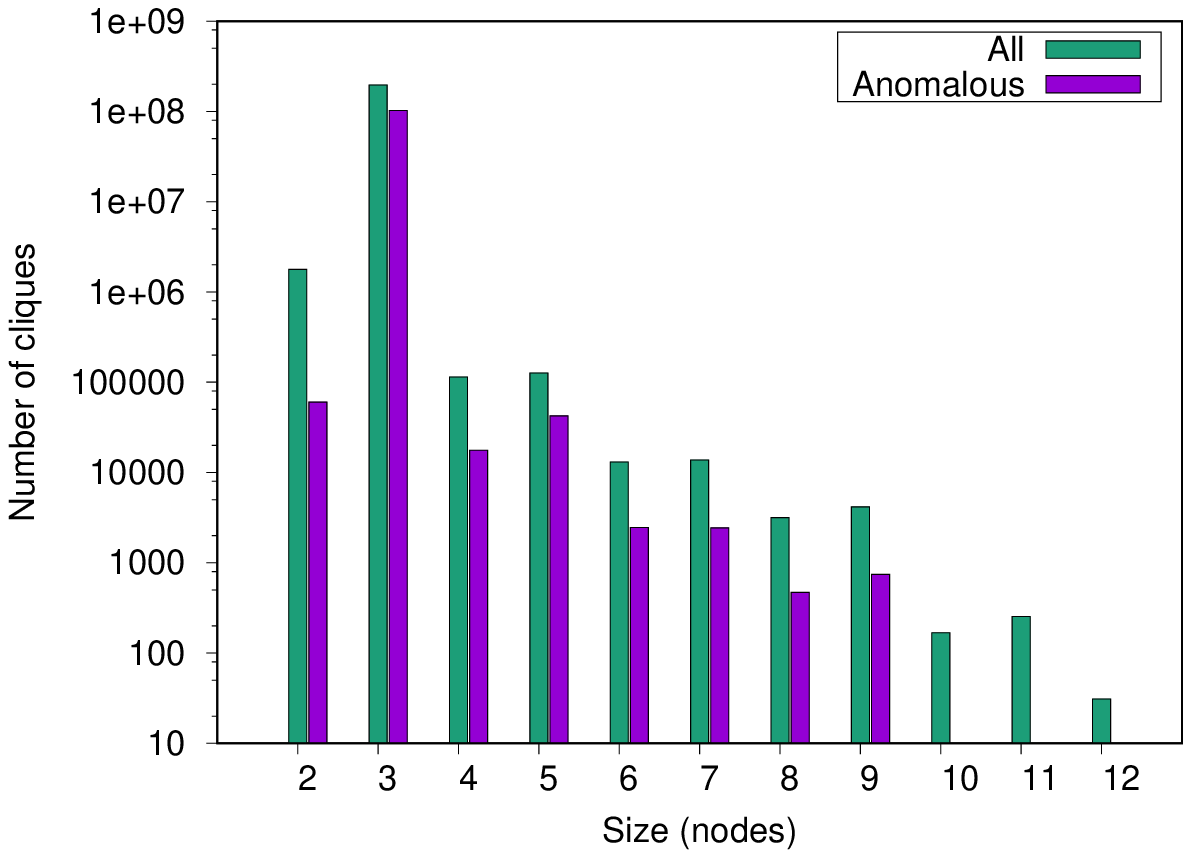}
\hfill
\includegraphics[width=0.45\columnwidth]{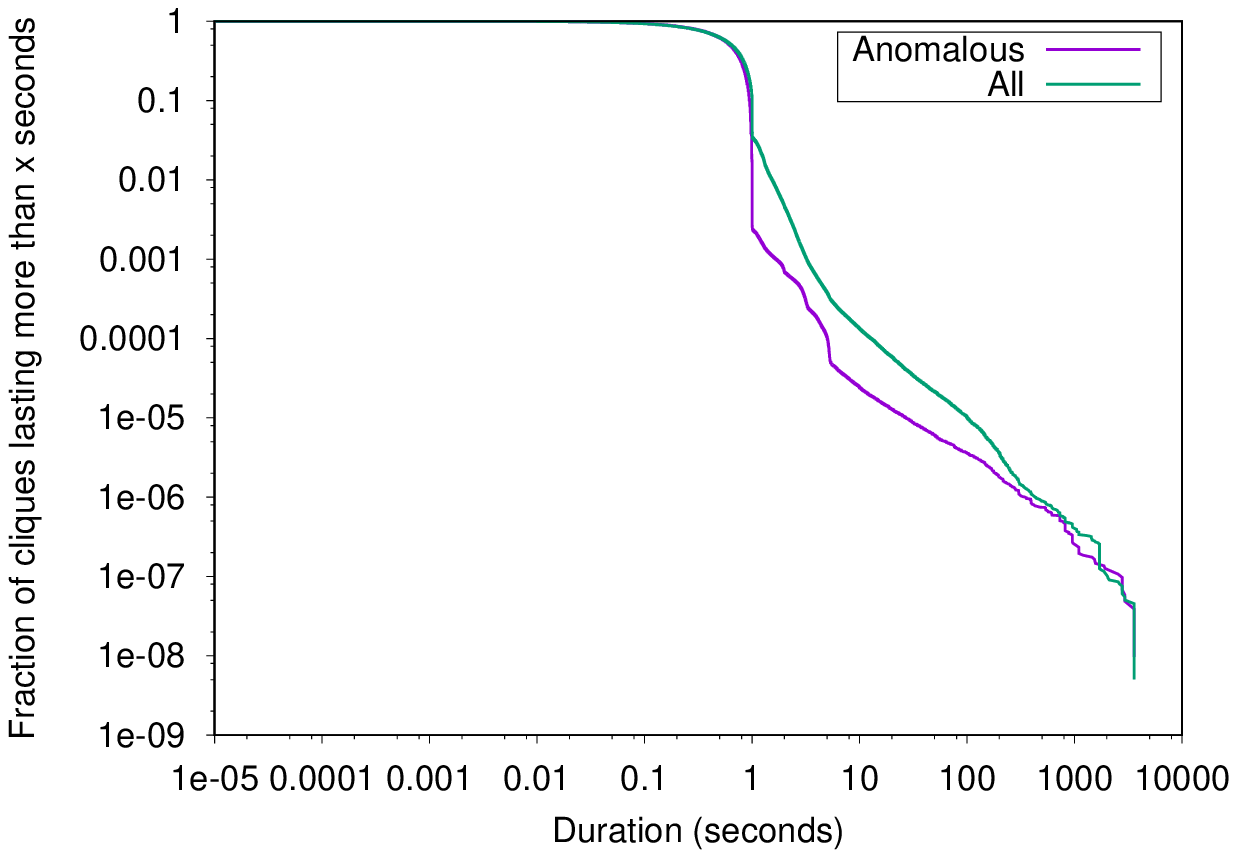}
\hfill\
\caption{
Size distribution (top) and duration inverse cumulative distribution (bottom) of the balanced cliques found by our sampling.
For instance, the fraction of cliques lasting more than $10$ seconds is $0.0001$.
}
\label{fig:sizes-dist}
\end{figure}

We show in Figure~\ref{fig:sizes-dist} the distribution of sampled clique sizes and the inverse cumulative distribution of their durations.  Many are of size $2$ or $3$, which has little interest: cliques of size $2$ are single links, and bipartite cliques of size $3$ are just composed of a node linked to two other nodes.  Since any node with $k$ neighbours leads to $k(k-1)/2$ balanced $3$-cliques, the large number of such cliques is unsurprising and brings no significant information. We therefore focus on cliques of size $4$ or more. Our sample contains $275,647$ such cliques.

The sampled cliques of $4$ nodes or more involve $29,744$ distinct nodes, $94$ of which anomalous. The fraction of anomalous nodes in these cliques therefore is $3.1\cdot 10^{-3}$, much larger than in the whole dataset, $5\cdot 10^{-4}$. This indicates that maximal balanced cliques are related to anomalous activity, as suspected.

\begin{figure}[!h]
\centering
\ \hfill
\includegraphics[width=0.45\columnwidth]{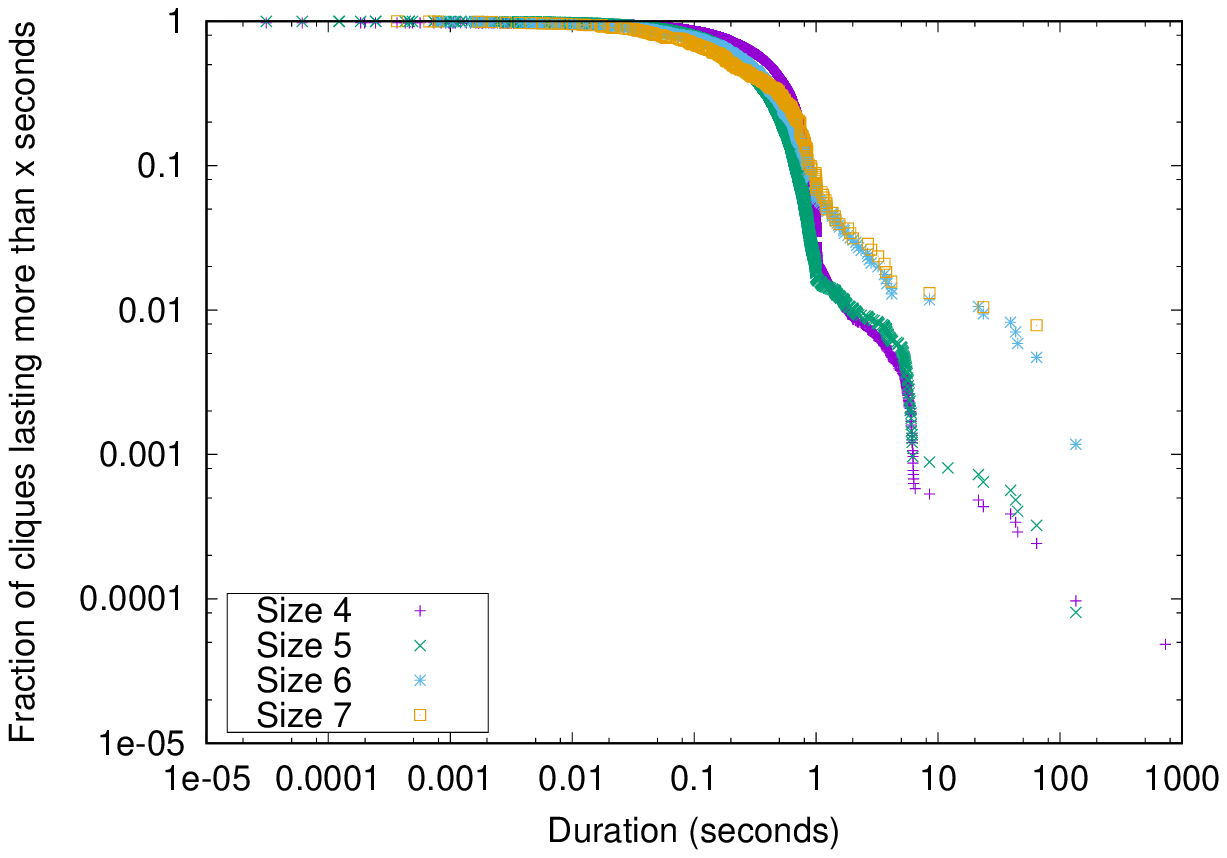}
\hfill
\includegraphics[width=0.45\columnwidth]{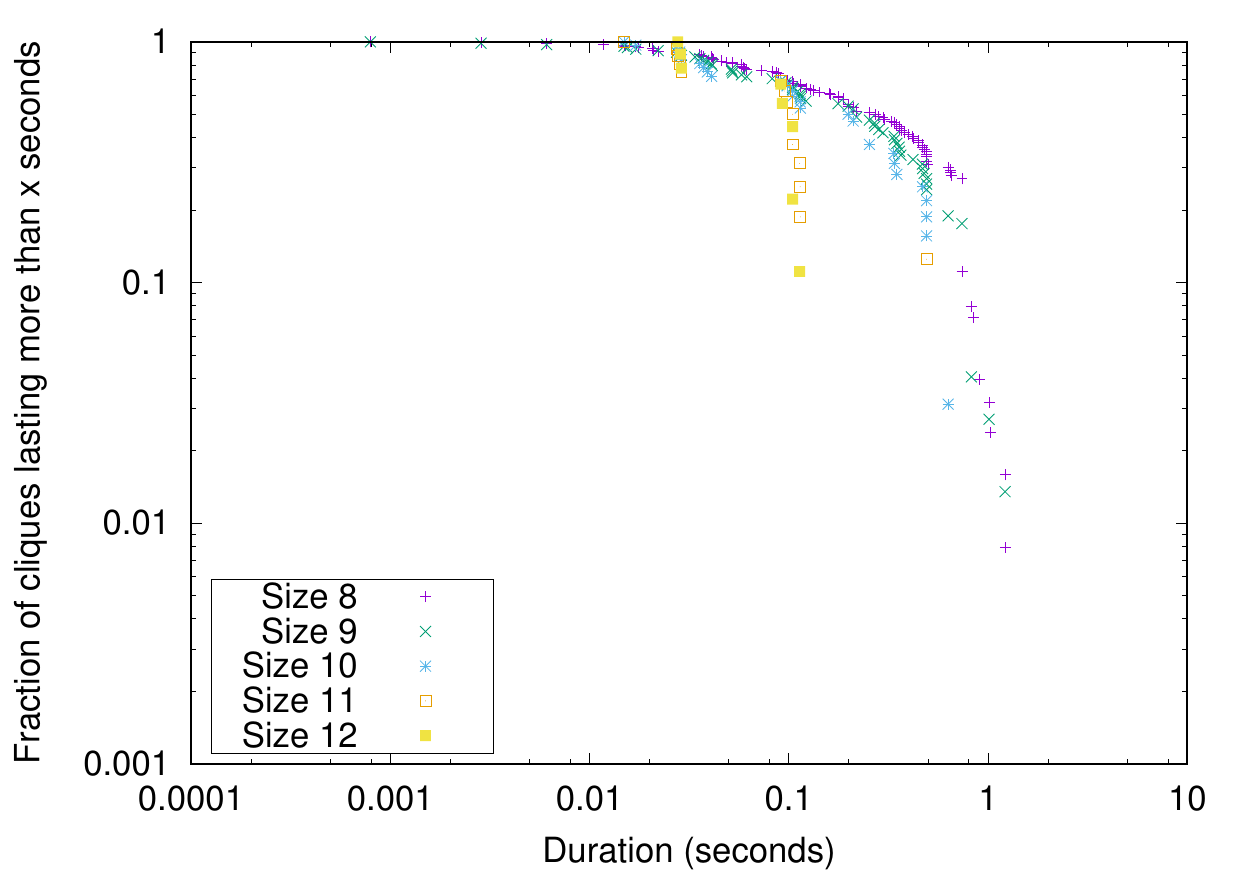}
\hfill\
\caption{
Inverse cumulative distribution of durations of sampled balanced cliques of a given size. Left: for sizes from 4 to 7 nodes. Right: for sizes from 8 to 12 nodes.
}
\label{fig:durations-ccdf-by-size}
\end{figure}

While most maximal balanced cliques have a duration close to $1$ second, the duration distributions show that there are very long cliques.
However, duration is highly influenced by size, as explained above, and so we display the duration distribution for each clique size separately in Figure~\ref{fig:durations-ccdf-by-size} (for readability, we show them in two plots).
As expected, the duration of large cliques is in general shorter than for cliques involving only few nodes. The $18$ cliques spanning more than $500$ seconds all have $4$ or $5$ nodes, the longest being of size $4$ and duration $1045.75$ seconds.
Still, there are cliques involving $7$ nodes for more than $10$ seconds, up to almost $100$ seconds. Larger cliques all have a duration close to $1$ second or less.
Importantly, among the $18$ cliques spanning more than $500$ seconds, $17$ are anomalous.

\begin{figure}[!h]
\centering
\ \hfill
\includegraphics[width=0.45\columnwidth]{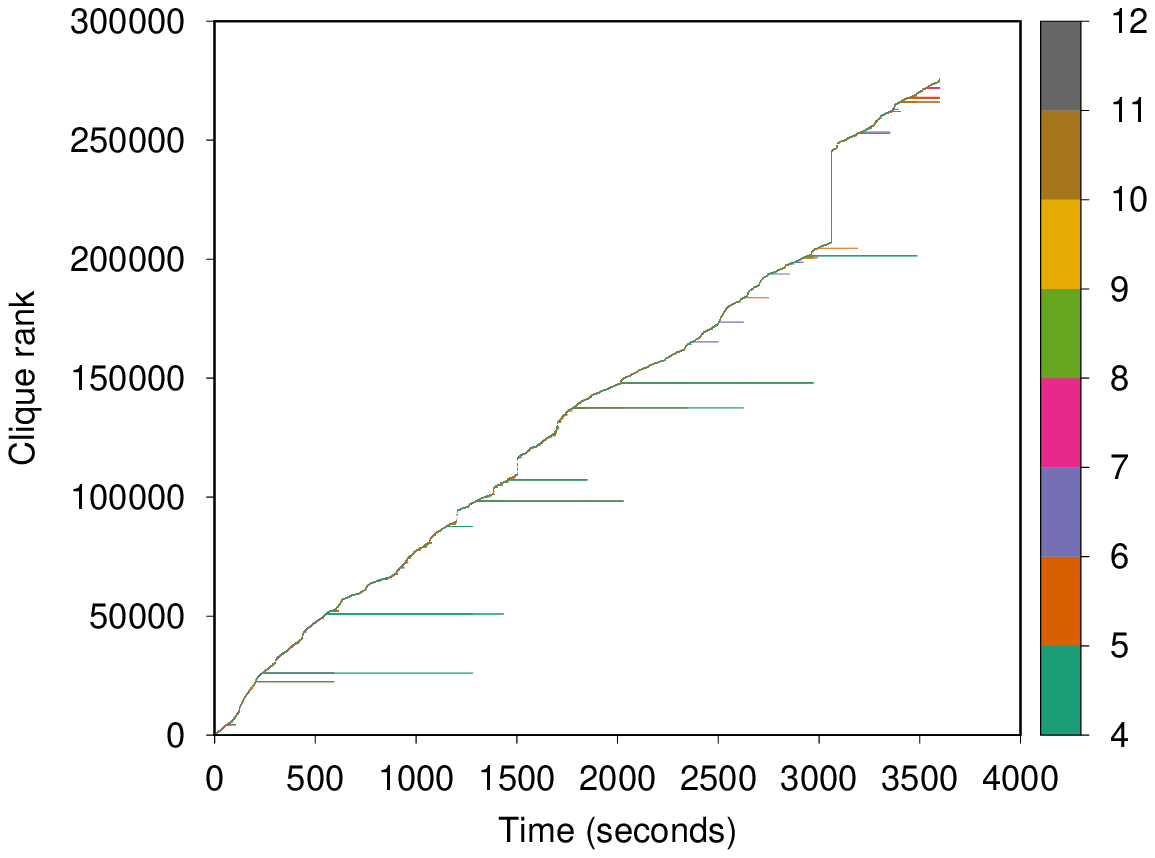}
\hfill
\includegraphics[width=0.45\columnwidth]{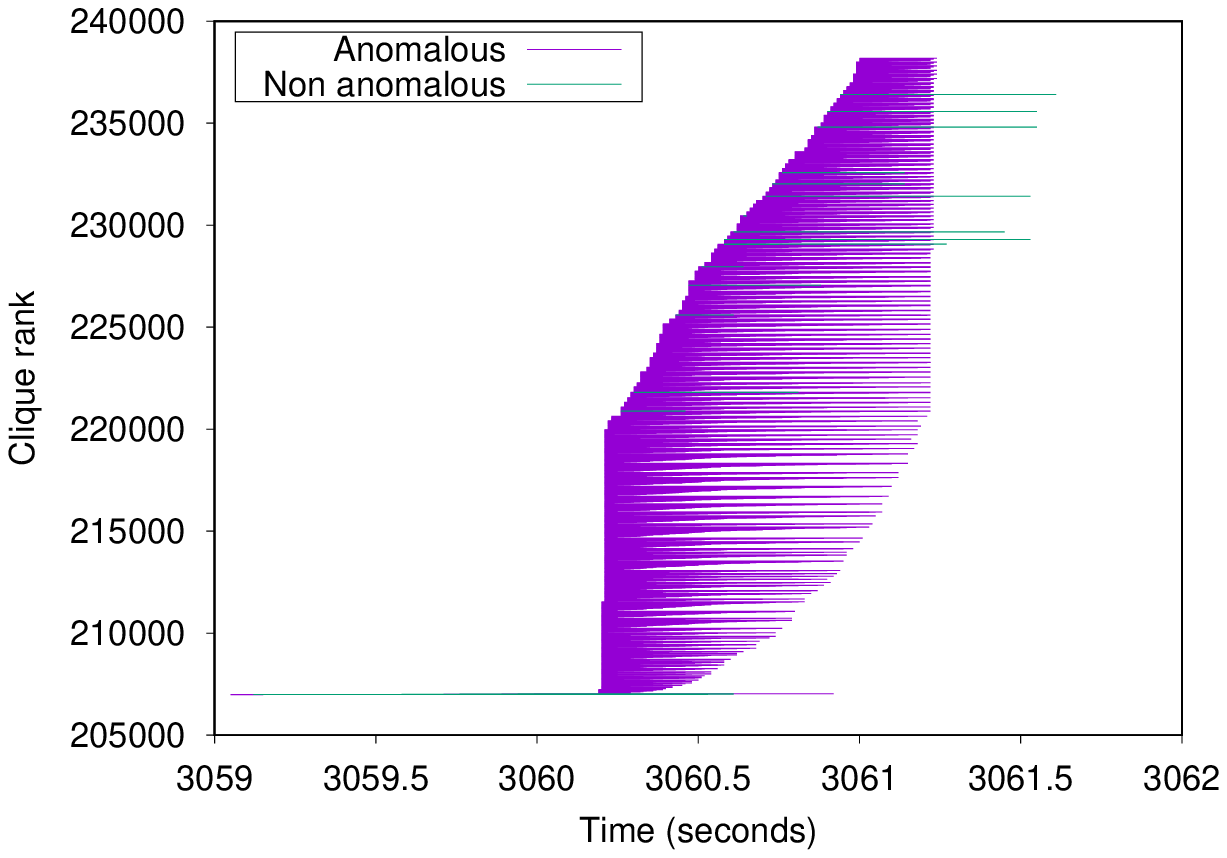}
\hfill\
\caption{
    {\bf Left:} Representation of the temporal features of our balanced cliques of size $4$ or more.
Each clique $(C_\top, C_\bot, I)$ is represented by a horizontal line of length $|I|$ with a color representing its size (color code on the right). These lines are ranked according to their beginning time.
{\bf Right:} Detailed view of the $[3059;3062]$ interval. A very large number of anomalous cliques start within the same second (in purple).}
  \label{fig:time-span}
\end{figure}

We deepen our understanding of large cliques by displaying in Figure~\ref{fig:time-span} the time span of all maximal balanced cliques of size $4$ or more: we sort cliques according to their starting time, which gives their vertical position in the plot. We then represent each clique $(C_\top,C_\bot,I)$ by an horizontal line from the beginning of $I$ to its end. Colors indicate clique size. This plot confirms our previous observations; it shows the prevalence of smaller cliques among the longest ones, and it shows that we succeed in finding significant cliques during the whole time span of the dataset.

However, it also displays a sharp increase shortly after time $3000$, corresponding to a large number ($31,201$) of short balanced cliques that start then. Figure~\ref{fig:time-span} right displays these cliques in more details, and shows that they almost all involve anomalous nodes. This confirms that clique structures are related to anomalous activity.

\begin{figure}[!h]
\centering
\ \hfill
\includegraphics[width=0.45\columnwidth]{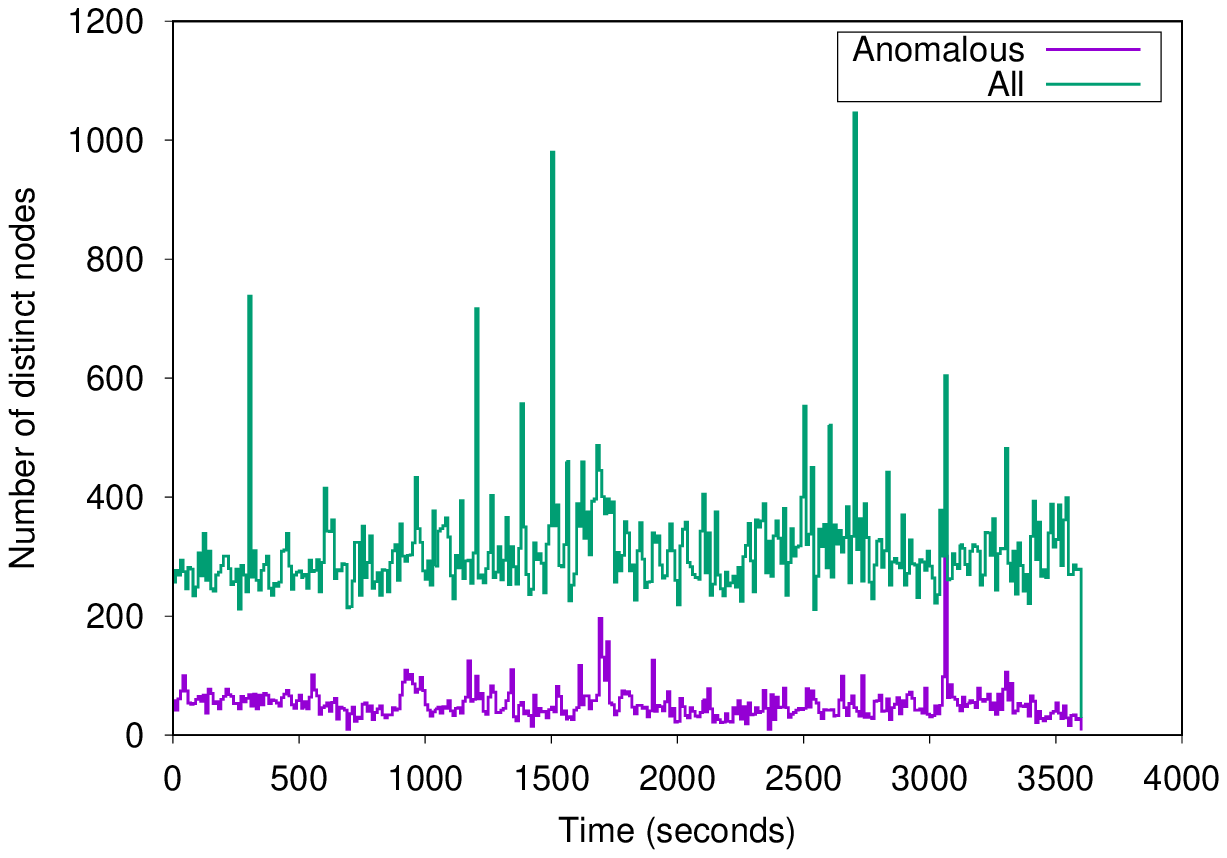}
\hfill
\includegraphics[width=0.45\columnwidth]{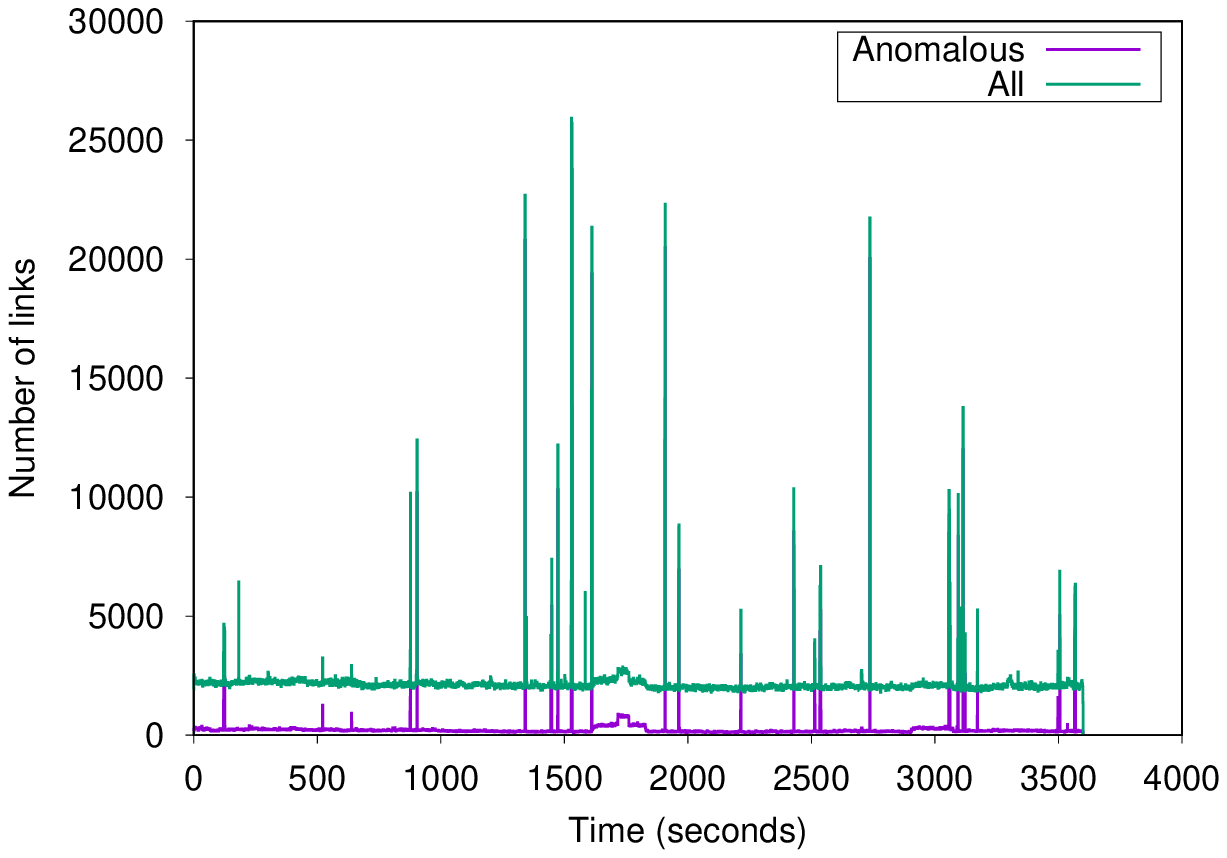}
\hfill\
\caption{Number of all and anomalous distinct nodes active in each second
  (left), and number of all and anomalous distinct links active in each second
  (right).}

\label{fig:per-second}
\end{figure}

Interestingly, it seems that these anomalies cannot easily be distinguished from other anomalies directly on simple plots like the number of distinct nodes or links over time, see Figure~\ref{fig:per-second}. Although there is a peak in both plots at time 3060, it is not different from other peaks, yet it is the only one corresponding to such a sharp increase in Figure~\ref{fig:time-span}. This indicates that cliques highlight specific features of this event, that are worth investigating further.

\begin{figure}[!ht]
\centering
  \includegraphics[clip,trim=0 6.9cm 0 6.9cm,width=0.45\columnwidth]{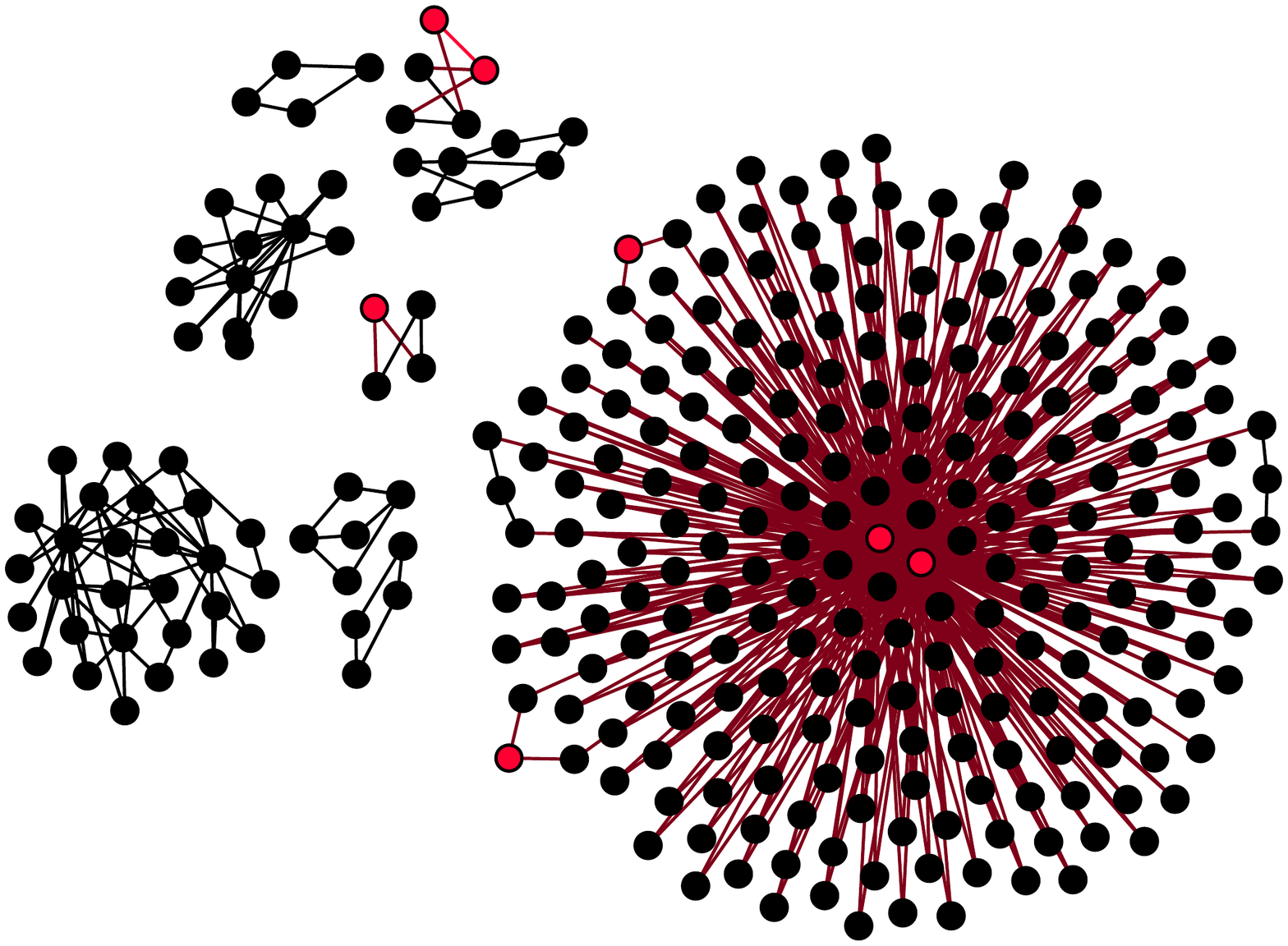}
\caption{The graph induced by all maximal balanced cliques starting between second $3059$ and $3061$, and involving at least $4$ nodes. Anomalous nodes known from \emph{MAWILab} are in red.}
\label{fig:gephi}
\end{figure}

We therefore display in Figure~\ref{fig:gephi} the graph induced by the balanced cliques we found at seconds $3059$, $3060$, or $3061$. The nodes of this graph are the nodes involved in at least one of these cliques, and they are linked together if this link occurs in at least one of these cliques. The graph has a large connected component with two anomalous nodes linked to $228$ distinct non-anomalous nodes, confirming that most of these cliques are the signature of a same event (they actually are parts of a much larger but unbalanced clique). This signature is typical of a coordinated scan in a class $C$ network.

\section{Related Work}
\label{sec-rw}

IP traffic has been extensively studied for decades with powerful approaches relying on signal processing and machine learning~\cite{DBLP:journals/ton/HimuraFCBAE13}, or graphs. In \cite{DBLP:journals/ijnm/AsaiFABE14}, the authors use graphs to represent temporal dependencies in traffic and characterize traffic behavior. In \cite{DBLP:journals/ton/XuWG14} the authors model the traffic as a bipartite graph, then used one-mode projections and clustering algorithms to discover behavioral clusters. Clustering communication behavior was also proposed in~\cite{DBLP:journals/jnw/JakalanJZQ15}, where authors discusses the relevant features before analysing significant nodes for long periods.

Graph-based approaches are however limited in their ability to capture temporal information, crucial for traffic analysis. Indeed, they generally rely on splitting data into time slices, and then aggregate traffic occurring into each slice into a (possibly weighted, directed, and/or bipartite) graph. One obtains this way a sequence of graphs, 
and one may study the evolution of their properties, see for instance~\cite{DBLP:journals/compnet/LatapyHM14}. 
However, choosing small time slices leads to almost empty graphs and bring little information. Conversely, large slices lead to important loss of information as the dynamics within each slice is ignored. As a consequence, choosing appropriate sizes for time slices is extremely difficult is a research topic in itself~\cite{DBLP:conf/conext/LeoCF15}. There is currently an important interdisciplinary effort for solving these issues by defining formalisms able to deal with both the structure and dynamics of such data. The link stream approach is one of them \cite{DBLP:journals/tcs/ViardLM16}, as well as temporal networks and time-varying graphs \cite{DBLP:conf/dsaa/WehmuthZF15,holme2015modern}. Up to our knowledge, these other approaches have not yet been applied to network traffic analysis.


\section{Discussion}
\label{sec-conclusion}

We have shown that cliques in bipartite link streams modeling of IP traffic allow to detect anomalous activity in IP traffic: long cliques of significant size involve anomalous nodes known from MAWILab, although they are rather small, and simultaneous apparition of many small cliques indicate coordinated activity like distributed scans. This work however only is a first step, and it raises many questions.

In particular, the computational cost of our method is prohibitive. Algorithmic work is therefore needed to design faster clique detection heuristics, and to search for quasi-cliques. One may also preprocess the stream by iteratively removing nodes of degree $1$, which represent a large fraction of the whole and cannot be involved in non-trivial cliques. Going further, one may use link streams to define many other structures of interest regarding anomalous traffic. Exploring other modeling assumptions is also appealing, in particular the fact that we linked nodes together if they exchanged packets at least every second (other time limits may be interesting), the fact that we considered undirected links, or the use of port or protocol information present in the data.

\bigskip
{\small \bf Acknowledgments.} {\small This work is funded in part by the European Commission H2020 FETPROACT 2016-2017 program under grant 732942 (ODYCCEUS), by the ANR (French National Agency of Research) under grants ANR-15-CE38-0001 (AlgoDiv) and ANR-13-CORD-0017-01 (CODDDE), and by the Ile-de-France program FUI21 under grant 16010629 (iTRAC).}


\bibliographystyle{splncs}
\bibliography{biblio}

\end{document}